\documentclass{webofc}

\usepackage[varg]{txfonts}   
\usepackage{hyperref}
\usepackage{url}
\RequirePackage{orcidlink}
\hypersetup{colorlinks=true,citecolor=blue,urlcolor=blue,linkcolor=blue}
\newcommand{\dquote}[1]{``#1''}
\newcommand{\aiinfn}{\ensuremath{\mathrm{AI\_INFN}}\xspace}
\newcommand\mlinfn{\ensuremath{\mathrm{ML\_INFN}}\xspace}
\newcommand\jhub{\ensuremath{\mathrm{JupyterHub}}\xspace}
\newcommand\jlab{\ensuremath{\mathrm{JupyterLab}}\xspace}
\newcommand\infncloud{\ensuremath{\mathrm{INFN\ Cloud}}\xspace}
\newcommand\datacloud{\ensuremath{\mathit{DataCloud}}\xspace}

\newcommand\ks{\ensuremath{\mathrm{Kubernetes}}\xspace}
\begin{document}

\title{Supporting the development of Machine Learning for 
fundamental science in a federated Cloud with the AI\_INFN platform}

\author{
    \firstname{Lucio} \lastname{Anderlini}\inst{1}\orcidlink{0000-0001-6808-2418}
    \and
    \firstname{Matteo} \lastname{Barbetti}\inst{2}\orcidlink{0000-0002-6704-6914}\fnsep\thanks{\email{matteo.barbetti@cnaf.infn.it}}
    \and
    \firstname{Giulio} \lastname{Bianchini}\inst{3,4}
    \and
    \firstname{Diego} \lastname{Ciangottini}\inst{3}\orcidlink{0000-0002-0843-4108}
    \and
    \firstname{Stefano} \lastname{Dal Pra}\inst{2}\orcidlink{0000-0002-1057-2307}
    \and
    \firstname{Diego} \lastname{Michelotto}\inst{2}\orcidlink{0000-0001-9909-8623}
    \and
    \firstname{Carmelo} \lastname{Pellegrino}\inst{2}\orcidlink{0000-0003-1530-4595}
    \and
    \firstname{Rosa} \lastname{Petrini}\inst{1}\orcidlink{0009-0003-3183-0429}\fnsep\thanks{\email{rosa.petrini@fi.infn.it}}
    \and
    \firstname{Alessandro} \lastname{Pascolini}\inst{2}
    \and
    \firstname{Daniele} \lastname{Spiga}\inst{3}\orcidlink{0000-0002-2991-6384}
}

\institute{
    Istituto Nazionale di Fisica Nucleare (INFN), Sezione di Firenze, Italy
    \and
    Istituto Nazionale di Fisica Nucleare (INFN), CNAF, Italy
    \and
    Istituto Nazionale di Fisica Nucleare (INFN), Sezione di Perugia, Italy
    \and
    Department of Physics, University of Perugia, Italy
}

\abstract{Machine Learning (ML) is driving a revolution in the way
scientists design, develop, and deploy data-intensive software. 
However, the adoption of ML presents new challenges for the computing 
infrastructure, particularly in terms of provisioning and orchestrating 
access to hardware accelerators for development, testing, and production.
The INFN-funded project \aiinfn (\dquote{Artificial Intelligence at INFN}) 
aims at fostering the adoption of ML techniques within INFN use cases 
by providing support on multiple aspects, including the provision of 
AI-tailored computing resources. It leverages cloud-native solutions in 
the context of INFN Cloud, to share hardware accelerators as effectively 
as possible, ensuring the diversity of the Institute's research 
activities is not compromised.
In this contribution, we provide an update on the commissioning of 
a \ks platform designed to ease the development of GPU-powered 
data analysis workflows and their scalability on heterogeneous, 
distributed computing resources, possibly federated as Virtual 
Kubelets with the interLink provider.}

\maketitle

\section{Introduction}
\label{sec:intro}

Over the past decade, \emph{Artificial Intelligence}~(AI) has seen 
rapid and widespread adoption, establishing itself as a standard tool 
for processing large, complex datasets, and extracting insights from 
multi-modal, multi-domain data~\cite{Bommasani2021FoundationModels}. 
The proliferation of text-to-image apps and the advent of AI-powered 
chat-bots have helped propel AI into the mainstream, consolidating its 
explosion in both usage and development.
Today, AI is reshaping the computing landscape, driving technological
evolution, influencing hardware market trends, 
and dominating software development worldwide.

Cloud computing has also played a key role in accelerating the
adoption of AI techniques by making computing power and data storage 
more accessible to developers by outsourcing resource management to 
service providers like AWS \cite{AWS} or GCP \cite{GCP}. 
This approach enables ready-to-use
software to be delivered directly to the end users, a model known as
\emph{Software-as-a-Service}~(SaaS). A typical example 
is JupyterHub \cite{JupyterHub}, 
which provides multiple users with access to notebooks for data 
visualization and interactive computation using resources
provisioned by the Cloud. The growing community of Data Scientists 
makes large use of this or similar technologies, contributing to 
expanding the catalog of libraries and tools available for developing 
and deploying AI applications.

The scientific community is closely following the evolution of Machine 
Learning~(ML), aiming at adapting advanced ML-based algorithms for 
fundamental research. This is particularly true in the High 
Energy Physics~(HEP) field, where researchers are exploring 
AI-driven computing solutions to accelerate the workflow, from data 
acquisition and simulation to physics analysis~\cite{Albertsson:2018maf}.
The paradigm of interactive computing is also gaining interest within 
the HEP community as a promising solution to meet the growing resource
demand expected from next-generation experiments. These efforts are
leading to the design and implementation of \emph{Analysis 
Facilities}~(AFs), a collection of infrastructures and services that 
integrates data, software, and computational resources to execute 
one or more elements of an analysis workflow~\cite{Ciangottini:2024vtl}.

The Italian National Institute for Nuclear Physics~(INFN) is the 
coordinating institution for nuclear, particle, astroparticle, medical 
and theoretical physics in Italy. It promotes, coordinates, and conducts
scientific research, along with the technological developments 
needed for the activities in these fields. In response to the 
paradigm shift driven by AI and Cloud technologies, INFN is reorganizing 
its computing infrastructure to support emerging trends, strengthening 
its resources~\cite{Grandi:2024fqt}, and expanding the range of services 
offered through INFN~Cloud\footnote{Cloud resources for INFN
research, \url{https://www.cloud.infn.it}.}~\cite{Salomoni:2024dft}.
Within this context, the \aiinfn (\dquote{Artificial 
Intelligence at INFN}) initiative was launched, aiming at sharing 
hardware resources, learning best practices, and developing 
AI-powered applications relevant to INFN research. 
In this document we present the \aiinfn platform and discuss 
its relevance to the outlined scopes.

\section{The AI\_INFN initiative}
\label{sec:ai-infn}

The \aiinfn initiative, 
aims to connect the scientific communities developing 
infrastructures, algorithms, and applications. 
It provides tools, computing infrastructures, and social connections to foster 
collaboration and facilitate the adoption of AI-powered computing technologies within 
INFN research fields. 

\aiinfn is organized in four work packages: procuring and operating computing 
infrastructure with hardware accelerators, organizing training events for Machine 
Learning adoption, building a community of ML experts and developers across INFN units, 
and developing the competence to profit from hardware accelerators beyond Graphics 
Processing Units (GPU), such as FPGAs and Quantum Processors, for AI model training
and inference.

The \aiinfn platform is a cloud-native toolset developed in collaboration with the 
\datacloud initiative operating \infncloud to support the activities of the four 
work packages of \aiinfn. In this document we present the \aiinfn platform and discuss 
its relevance to the outlined scopes.

The \aiinfn platform is a SaaS provided by \infncloud. The underlying specialized and 
dedicated hardware was designed for High Performance Computing (HPC) tasks and is hosted 
and managed at INFN CNAF in Bologna, few steps away from the pre-exascale CINECA
supercomputer Leonardo\footnote{More details available at \url{https://leonardo-supercomputer.cineca.eu}}.
The infrastructure comprises four servers, clustered in an 
OpenStack \cite{OpenStack} tenancy, acquired 
and installed between 2020 and 2024 as the demand for Cloud-accessible HPC computing 
resources increased:
\begin{itemize}
    \item \textbf{\emph{Server 1}} (2020), with 
        64 CPU cores, 
        750 GB of memory, 
        12 TB of NVMe disk, 
        eight NVIDIA Tesla T4 GPUs and 
        five NVIDIA RTX 5000 GPUs;
    \item \textbf{\emph{Server 2}} (2021), with
        128 CPU cores, 
        1024 GB of memory, 
        12 TB of NVMe disk, 
        two NVIDIA Ampere A100 GPU,
        one NVIDIA Ampere A30 GPU,
        two AMD-Xilinx U50 boards and 
        an AMD-Xilinx U250 board;
    \item \textbf{\emph{Server 3}} (2023), with
        128 CPU cores, 
        1024 GB of memory, 
        24 TB of NVMe disk, 
        three NVIDIA Ampere A100 GPUs and
        five AMD-Xilinx U250 boards;
    \item \textbf{\emph{Server 4}} (2024), with
        128 CPU cores, 
        1024 GB of memory, 
        12 TB of NVMe disk, 
        one NVIDIA RTX 5000 GPUs and
        two AMD-Xilinx Versal V70 boards.
\end{itemize}
%

Before \aiinfn started its activity in January 2024, the farm was maintained by 
another INFN initiative named \mlinfn that, created as a \emph{proof-of-concept} for
designing a platform for sharing accelerated resources, was developed with a provisioning 
model relying on Virtual Machines (VMs) assigned to groups of users developing a data analysis 
or Machine Learning study~\cite{Anderlini:2024oup}.
During the late period of \mlinfn, however, an increase in the user base highlighted 
some limitations to the efficiency of this provisioning model. These limitations were
related to administrative and user-support burden, very long idling times, and 
dangerous eviction of the stateful user's deployments.
In 2023, the security risks grew to an unacceptable level and called for the introduction 
of an alternative model enabling users to tune the resources provisioned to their
cloud-based computing environment, without becoming administrators of a multi-user 
web service. 

At the time of writing, 72 researchers working on 16 research activities have requested 
and gained access to the platform. 
On average, 10 to 15 researchers connect at least once to the platform in a working day.
Two dedicated clones of the \aiinfn platform were temporarily deployed at CNAF and at ReCaS Bari 
to provision the GPU-accelerated resources to the 30 participants of the 
first AI\_INFN hackathon, an advanced training event organized in Padua in November 2024\footnote{More details available at \url{https://agenda.infn.it/event/43129}.}.

\section{SaaS provisioning model, the AI\_INFN platform}
\label{sec:ai-infn-arch}

The \aiinfn platform was deployed on a \ks \cite{Kubernetes}
cluster spanning on at least three VMs within the dedicated OpenStack tenancy providing 
part of the storage resource, the monitoring infrastructure and the \ks 
control plane. A minimal amount of compute resources is also provisioned 
to make it possible for users to access their data on the platform 
anytime~\cite{Anderlini:wmlq2024}.
Additional compute resource provided by VMs can be attached to the cluster 
and detached to be used as standalone machines running an 
Ansible \cite{Ansible} playbook, or reassigned 
to another cluster in the same tenancy. \aiinfn users are identified through 
INFN Cloud Indigo IAM~\cite{indigoiam} instance. Once authenticated, users can 
configure and spawn their \jlab instance using \jhub.


The main platform file system is distributed through the containers via
NFS. One of the platform nodes runs an NFS server in a \ks pod and 
exports data to the containers spawned by \jhub.
At spawn time, \jhub is configured to create the user's home directories 
and project-dedicated shared volumes. 
A special directory of the platform file system, that users can 
use directly or clone and extend in their directories, is reserved for 
distributing managed software environments, configure using virtual environments.
The platform file system is subject to regular encrypted backup. 
Backup data is stored in a remote Ceph~\cite{ceph} volume provisioned 
by \infncloud using the \emph{BorgBackup} \cite{BorgBackup} package to ensure data deduplication.

Large datasets must be stored in a centralized object storage service
based on Rados Gateway \cite{RadosGateway} and centrally managed by \datacloud.
To ease accessing the datasets with the Python frameworks commonly 
adopted in Machine Learning projects, a patched version of \texttt{rclone} \cite{rclone} 
was developed to enable mounting the user's bucket
in the \jlab instance using the same authentication token used to
access \jhub. The mount operation is automated at spawn time.

To address the bandwidth limitations of a virtual file system with a
remote backend, which can hinder iterative training and data analysis 
requiring to process the whole dataset multiple times, the \aiinfn 
platform provides also an ephemeral file system. This system is 
mapped directly to a logical volume on the hypervisor's NVMe storage. The indications
for the users is to copy the required data to this fast volume at the beginning of each session. 
These ephemeral volumes are also useful as a cache for intermediate results or 
to extend RAM through memory mapping.

At the opposite extreme of the I/O performance spectrum there are distributed virtual file systems
that can be mounted on multiple computing resources, enabling the sharing of notebooks and user-defined 
computing environments across multiple computing sites and compute backends.
JuiceFS \cite{JuiceFS} is a cloud-based, high-performance, 
POSIX-compliant distributed file system specifically designed 
for multi-cloud and serverless computing. 
It decouples data and metadata delegating these tasks
to highly optimized third party projects, combining a metadata engine 
implemented with either key-value databases (such as Redis \cite{redis-link}) 
or relational database management systems (such as PostgreSQL \cite{PostgreSQL}) with storage 
systems accessed through S3, WebDAV or other high-throughput protocols.

Finally users can install or upgrade packages in their containers.
Installing new software will introduce ephemeral modifications in 
OverlayFS layer on top of the container file system. 
A more effective and popular alternative to installing packages 
in the container is to rely on the binaries distributed through 
the CERN VM file system (\texttt{cvmfs}) \cite{Buncic:2010zz}. 
CVMFS, used to distribute software through the nodes of the 
WLCG, is made available to the platform users through a 
\ks installation that shares the caches among different 
users and sessions.

In the \aiinfn platform, the NVIDIA GPU operator \cite{GpuOperator}, 
is used to install and maintain the GPU drivers.
The NVIDIA GPU Operator on \ks streamlines the management and deployment of 
NVIDIA GPU resources by automating the installation and configuration of 
required components within a \ks cluster.
In general, the GPU Operator is designed to simplify the management of 
clusters with a large number of GPU accelerators, enabling efficient scaling 
and resource optimization for GPU-accelerated workloads such as AI, ML, and data processing.
This approach enables centralized management of GPUs, ensuring a consistent and 
scalable configuration across all nodes in the cluster, while simplifying maintenance
and updates.

One of the most common support requests with the VM-based provisioning model 
involved setting up a GPU-accelerated Python software stack. While the TOSCA
template~\cite{TOSCA} and Ansible playbook handled the installation of the NVIDIA driver 
and runtime, choosing and installing the Python libraries required for the 
application was left to the users.
Typically, managing a data science project's dependencies is the responsibility 
of developers and analysts, and many analysis projects require multiple computing
environments. When introducing the \aiinfn platform, particular attention was given to ensuring user sessions were highly customizable and adaptable by providing mechanisms for users to create and manage their own computing environments. The most radical customization option is to build and pick a custom OCI image. Both communities and individual users can modify the default OCI image by adding system libraries or software packages or by altering the \jlab service itself.
While users often prefer conda \cite{conda} for custom software environments, Apptainer \cite{Apptainer} images are gaining popularity. Unlike conda, which consists of thousands of small files, Apptainer uses SquashFS \cite{Squashfs}, a compressed read-only file system, to package the entire environment into a single file. This makes Apptainer images easier to share and distribute through object stores.
The \aiinfn platform provides documentation to help users export conda environments as Apptainer images and use them as Jupyter kernels. It also offers pre-built conda environments and Apptainer images with software versions optimized for GPU-accelerated Machine Learning frameworks. Users can clone these environments and add project-specific dependencies, typically related to data loading and visualization, and independent of the GPU 
software stack.
A notable exception is represented by the software environment to develop
Quantum Machine Learning (QML), featuring Python modules that simulate 
the effect of quantum operators on GPU and therefore requiring the same
attention as other GPU-accelerated ML libraries to match the versions 
of the underlying software.
In addition, Apptainer images specialized for the data processing of the LHC 
experiments can be obtained via CVMFS.

A dedicated monitoring and accounting system has been set 
up for the platform in order to effectively control the use of all the 
platform's resources and in particular of the GPUs. 
Several metric exporters have been configured to collect the information 
of interest and then expose it to a Prometheus \cite{Prometheus-link}
instance 
running in the platform. 
Some of these exporters are already available as Free and Open Source Software,
such as Kube Eagle \cite{KubeEagle}, which manages information 
about the use of the cluster's CPUs and memory resources by the various 
components of \ks, or NVIDIA GPU DCGM exporter \cite{DCGM}.
Other exporters were developed on purpose, for example to monitor the usage
of storage resources.
All the metrics collected by Prometheus are then made visible and accessible 
through a Grafana \cite{grafana-docs} dashboard. 
Grafana is run in a VM independent of the platform cluster and is used to 
monitor other VMs in the \aiinfn OpenStack tenancy. 
It also hosts a PostgreSQL database for the accounting metrics, 
updated at regular intervals by averaging the metrics obtained from the 
monitoring Prometheus service.

\section{Offloading: scale the applications beyond cluster boundaries}
\label{sec:offloading}

While the \aiinfn platform is primarily conceived as the to-go solution for 
researchers moving their first steps with hardware-accelerated and machine 
learning software development, it was designed to enable packaging and scaling
the developed applications with computing resources made 
available in remote computing centers with a mechanism known as 
\emph{offloading}.

The architecture of the offloading capabilities of the \aiinfn platform
consists of few self-consistent components that \emph{can} be used by 
the application to scale beyond the single notebook instance.
Indeed, different components may introduce different limits or limitations 
on the scaling capabilities and should be selected wisely on a per-application
basis.

Users are allowed to scale beyond their notebook instance by creating \ks 
jobs, enqueued and assigned to either local or remote resources by the 
Kueue controller \cite{Kueue}. 
Kueue is designed to use local resources in an 
opportunistic way, configuring the running batch jobs to be immediately evicted 
in case new notebook instances are spawned pushing the cluster in a condition of 
resource contention. User do not create jobs directly accessing \ks APIs, 
but passing through a dedicated microservice, named \emph{vkd}, that validates 
user's request based on memberships criteria and manage \ks secrets that 
are not intended to be exposed to users, but still are needed for their jobs to 
be executed in the platform.

An interesting feature supported by \emph{vkd} is the ability of cloning the notebook
instance, replacing the start-up commands spawning the notebook with user-defined 
commands. In practice, these \emph{Bunshin Jobs} provide a very simple interface to 
scale the application within the cluster boundaries as the applications developed 
within the notebook instance are guaranteed to run identically in the cloned instances. 

\begin{figure}[t!]
\centering
    \begin{minipage}{0.43\textwidth}
        \includegraphics[width=1.03\textwidth, clip]{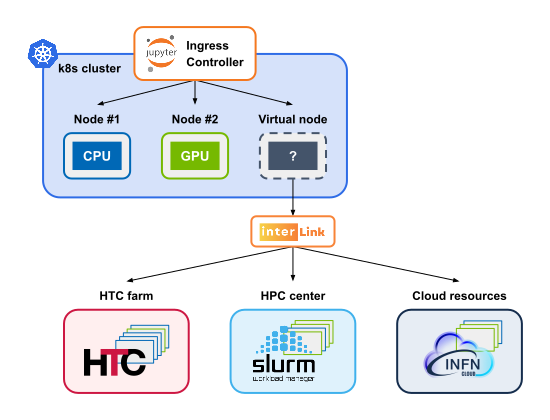}
        \caption{
            \label{fig:offloading}
            The \aiinfn platform enables offloading using InterLink as a virtual Kubelet provider to offload workloads from a Kubernetes cluster to external resources. InterLink interfaces with HTCondor, Slurm, and Podman via dedicated plugins, providing transparent access to HPC and cloud infrastructures.
            }
    \end{minipage}
    \hfill
    \begin{minipage}{0.48\textwidth}
        \includegraphics[width=\textwidth, clip]{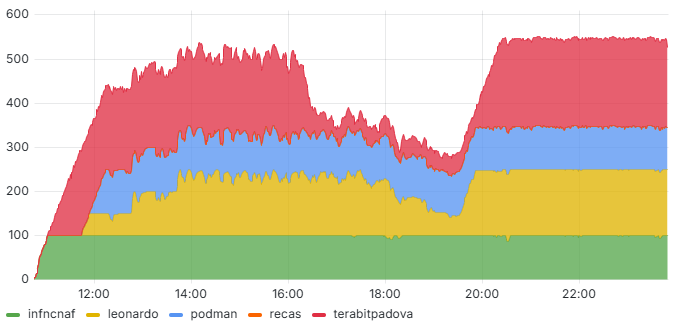}
        \caption{
            \label{fig:testfeb}
            Scalability test involving:  
                the INFN-Tier1 at CNAF in Bologna provisioned via HTCondor (labeled \texttt{infncnaf});
                the CINECA Leonardo super computer in Bologna provisioned via Slurm (\texttt{leonardo});
                a Virtual Machine in the Cloud provisioned via Podman (\texttt{podman});
                a Terabit \emph{HPC-Bubble} in Padova provisioned via Slurm (\texttt{terabitpadova}).
                The label \texttt{recas} in the legend refers to a WLCG Tier-2 site in Bari, integrated, but not taking part to the test.
            }
    \end{minipage}
\end{figure}

Users can flag their jobs at submission time as \emph{compatible with offloading}. 
The compatibility of a job with offloading should be evaluated considering technical aspects
(for example, an offloaded job cannot rely on the local storage resources such as NFS), 
practical consideration (for example, the longer delay between submission and execution in 
large data centers may make offloading ineffective for very short jobs), and policy
restrictions (for example, secrets to access confidential data cannot be shared with a 
remote data center). Kueue may then assign jobs marked as \emph{compatible with offloading} 
to \emph{virtual nodes}.

Virtual nodes are \ks nodes that are not backed by a Linux kernel but mimic a 
\ks \emph{kubelet} in the interactions with the \ks API server. 
The software component providing this interface is named \emph{Virtual Kubelet} \cite{vk} and 
is designed to ease the integration with various resource providers. The \aiinfn 
platform relies on the InterLink~\cite{Ciangottini:chep2024} provider. A further abstraction layer 
defining a simplified set of REST APIs that can be implemented by the so-called 
\emph{InterLink plugins} providing the actual access to the compute resources. 
At the time of writing, the \aiinfn platform is interfaced with plugins accessing 
HTCondor~\cite{htcondor}, Slurm~\cite{slurm} and Podman \cite{podman} resources. 
Following a recent integration test, a \ks plugin will be brought to production soon.
A schematic representation of the architecture is provided in Figure~\ref{fig:offloading}.

To ease the deployment of application in the remote data centers, the \aiinfn platform
relies on dedicated and distributed file system based on JuiceFS using
Redis as metadata engine and an S3 endpoint for data storage. 
The secrets to mount the shared file system are shipped to the remote 
data center that, if allowed by site-specific policies, can make it available to the 
applications as a FUSE file system. Relying on the distributed file system drastically
hinder the scalability of the developed application, but provides a precious intermediate
level between cluster-local development and multi-site distributed production. 

Figure~\ref{fig:testfeb} reports a recent scalability test involving resources provisioned 
by four different sites, without distributing the file system and for CPU-only payloads
of the LHCb Flash Simulation~\cite{Barbetti:2906203}.
The plot illustrates the increase in job counts and their 
distribution across multiple computing sites. Stable contributions 
were maintained throughout the test by the INFN-Tier1 at CNAF in 
Bologna, provisioned via HTCondor. Comparable stability was 
observed from a Podman-based VM, whereas CINECA’s Leonardo 
supercomputer, provisioned via Slurm, showed greater 
variability in job activity 
during the initial hours. Later in the test, the Terabit HPC-Bubble, also provisioned via Slurm, was integrated 
and demonstrated consistent, sustained usage. The pronounced increase 
in job activity during the evening suggests improved scheduling 
efficiency. Overall, the test confirms the effectiveness of 
integrating multiple sites with heterogeneous provisioning mechanisms.

\section{Conclusion}
\label{sec:conclusion}


Machine Learning and Artificial Intelligence have been reshaping the landscape 
of data processing and data analysis applications, making it easier for 
data analysts and data scientist to accelerate a variety of computing-intensive
tasks on GPUs.
The \aiinfn initiative is developing and serving a highly customizable development 
platform, integrated in the service portfolio of \infncloud 
and provisioning different GPU models, possibly installed in remote computing centers 
through offloading techniques. 
%
Although primarily a research and development project, the \aiinfn platform is 
gaining recognition from several small experiments within the \aiinfn research 
lines as a potential provider of GPU-accelerated computing resources, and is
poised to play a trailblazing role in the future landscape of the INFN 
computing infrastructure.

\section*{Acknowledgements}
The work presented in this paper is performed in the framework 
of Spoke~0 and Spoke~2 of the ICSC project -- \emph{Centro Nazionale 
di Ricerca in High Performance Computing, Big Data and Quantum 
Computing}, funded by the NextGenerationEU European initiative through 
the Italian Ministry of University and Research, PNRR Mission~4, 
Component~2: Investment~1.4, Project code CN00000013 - CUP 
I53C21000340006.

\bibliography{main}

\end{document}